\documentclass[prx,aps,floatfix,longbibliography,superscriptaddress,twocolumn,final,notitlepage]{revtex4-2}
\RequirePackage{amsmath}
\RequirePackage{hyperref} 
\RequirePackage{amssymb}
\RequirePackage{graphicx}
\RequirePackage{bbm}
\RequirePackage{subfigure}
\RequirePackage{xcolor}
\usepackage{natbib}
\usepackage[nameinlink,capitalize]{cleveref}

\setcitestyle{round}

\setcitestyle{round}

\setcitestyle{round}

\newcommand{\be}{\begin{equation}}
\newcommand{\ee}{\end{equation}}
\newcommand{\bea}{\begin{eqnarray}}
\newcommand{\eea}{\end{eqnarray}}

\newcommand{\ccup}[1]{\left\{#1\right\}}

\usepackage{float}
\makeatletter
\let\newfloat\newfloat@ltx
\makeatother
\usepackage{algorithm}
\usepackage{algpseudocode}
\usepackage{amsmath,fixmath}
\usepackage{fixmath}
\usepackage{setspace}
\usepackage{float}

\usepackage{booktabs}
\usepackage{adjustbox}

 \usepackage{multirow}
\definecolor{shadecolor}{gray}{0.9}
\usepackage{tabu}

\usepackage[normalem]{ulem} 
\usepackage{soul}

\newcommand{\hymmsbm}{\mbox{\small Hy-MMSBM}}
\newcommand{\hycosbm}{\mbox{\small HyCoSBM}}
\newcommand{\repolink}{\href{https://github.com/badalyananna/HyCoSBM}{github.com/badalyananna/HyCoSBM}}
\newcommand{\clique}{\mbox{{\small Clique-Exp}}}

\graphicspath{{figures}}
\begin{document}

\title{Structure and inference in hypergraphs with node attributes}

\author{Anna Badalyan}
\affiliation{Max Planck Institute for Intelligent Systems{$\text{,}$} Cyber Valley{$\text{,}$} T{\"u}bingen 72076{$\text{,}$} Germany}

\author{Nicol{\`o} Ruggeri}
\affiliation{Max Planck Institute for Intelligent Systems{$\text{,}$} Cyber Valley{$\text{,}$} T{\"u}bingen 72076{$\text{,}$} Germany}
\affiliation{Department of Computer Science{$\text{,}$} ETH{$\text{,}$} Z{\"u}rich 8004{$\text{,}$} Switzerland}

\author{Caterina De Bacco}
\email{caterina.debacco@tuebingen.mpg.de}
\affiliation{Max Planck Institute for Intelligent Systems{$\text{,}$} Cyber Valley{$\text{,}$} T{\"u}bingen 72076{$\text{,}$} Germany}

\begin{abstract}
Many networked datasets with units interacting in groups of two or more, encoded with hypergraphs, are accompanied by extra information about nodes, such as the role of an individual in a workplace. Here we show how these node attributes can be used to improve our understanding of the structure resulting from higher-order interactions. We consider the problem of community detection in hypergraphs and develop a principled model that combines higher-order interactions and node attributes to better represent the observed interactions and to detect communities more accurately than using either of these types of information alone.
The method learns automatically from the input data the extent to which structure and attributes contribute to explain the data, down weighing or discarding attributes if not informative. Our algorithmic implementation is efficient and scales to large hypergraphs and interactions of large numbers of units. We apply our method to a variety of systems, showing strong performance in hyperedge prediction tasks and in selecting
community divisions that correlate with attributes when these are informative, but discarding them otherwise. Our approach illustrates the advantage of using informative node attributes when available with higher-order data.
\end{abstract}

\maketitle


Over recent years, systems where units interact in groups of two or more have been increasingly investigated. Such higher-order interactions  have been observed in a wide variety of systems, including cellular networks \cite{klamt2009hypergraphs}, drug recombination \cite{zimmer2016prediction}, ecological communities \cite{mayfield2017higher} and functional mapping of the human brain \cite{giusti2016two}.

These systems can be better described by hypergraphs, where hyperedges encode interactions among an arbitrary number of units \cite{battiston2020networks, battiston2021physics}. 
Often, research in this area solely considers the topology of hypergraphs, that is, a set of nodes and their higher-order interactions. 
Many hypergraph datasets, however, include attributes that describe properties of nodes, such as the age of an individual, their job title in the context of workplace interactions, or the political affiliation of a voter.
In this work, we consider how to extend the analysis of hypergraphs to incorporate this extra information. 

We focus on the relevant task of community detection, where the goal is to cluster nodes in a hypergraph. Community detection algorithms solely based on interactions tend to cluster nodes based on notions of affinity between communities, cluster separation, or other arguments similar to those classically utilized on graphs \cite{fortunato2010community}. However, one can assume that relevant information about the communities and the hyperedge formation mechanism is additionally contained in the attributes accompanying a dataset.

For instance, students in a school have been observed to interact more likely in groups that involve individuals in the same classes \cite{mastrandrea2015contact}. A similar observation was also made for dyadic networks, where incorporating node attributes helped in community detection and other related inference tasks, e.g. prediction of missing information \cite{contisciani2020community,yang2013community,fajardo2022node,newman2016structure, tallberg2004bayesian}.

Several tools have been developed for community detection in higher-order data \cite{eriksson2021choosing,carletti2021random,vazquez2009finding,zhou2006learning}. Methods based on statistical inference have established themselves as effective tools in this direction, as they are both mathematically principled and have a high computational efficiency \cite{contisciani2022inference,ruggeri2023generalized,chodrow2021generative}.

Here, we build on these approaches to incorporate node attributes into a community detection framework for higher-order interactions. More precisely, we follow the principles behind generative models for networks, which incorporate community structure by means of latent variables that are inferred directly from the observed interactions~\cite{ball2011efficient, debacco2017community,goldenberg2010} and extend them to incorporate extra information on nodes. 

The model we propose has several desirable features. It is flexible, as it can be applied to both weighted and unweighted hypergraphs, it can incorporate different node attributes, categorical or binary, and it outputs overlapping communities, where nodes can belong to multiple groups simultaneously. Furthermore, the model does not assume any a priori correlation structure between the attributes and the communities. Rather, it infers such a connection directly from the data. The extent of this contribution can vary based on the dataset. In the favorable case where attributes are correlated well with the communities, our model exploits such additional information to improve community detection. This is particularly beneficial in situations where data is sparse or when data availability is limited to an incomplete set of observations. In less favorable situations where correlation is low (for instance when the attributes do not align  with the mechanism generating higher-order interactions), the model can nevertheless either discard or downweigh this information. 

In some cases, a system can be explained well by different community divisions. Our model allows selecting a particular community structure guided by the desired attribute, provided that it is informative,  as measured automatically by fitting the data. This allows a practitioner to focus the analysis of group interactions on some particular node characteristic.

Finally, our model is computationally efficient, as it scales to large hypergraphs and large hyperedge sizes. This feature is particularly relevant in the presence of higher-order interaction, where the increased computational complexity limits the range of models that can be practically implemented into viable algorithms. 

Few works are available that investigate community detection in hypergraphs in presence of node attributes \cite{li2023efficient,fanseu2021hypergraph,du2019hybrid}, but they are limited to clustering nodes without providing additional probabilistic estimates. Furthermore, they can be computationally burdening, or they typically rely on stronger assumptions about the nature of the data (e.g. assume real-valued weights) or the communities (e.g. nodes can only belong to one group).

\section*{Results}
\subsection*{The Model}
\label{sec: model}
We propose a probabilistic model that incorporates both the structure of a hypergraph, i.e. the interactions observed in the data, and additional attributes (or covariates) on the nodes. These two types of information, which we call structural and attribute information, have been previously shown to be informative in modeling community structure in networks, when there is correlation to be exploited\cite{contisciani2020community,yang2013community,fajardo2022node,newman2016structure}.

We denote a hypergraph as $H=(V, E, A)$, where $V=\{1, \ldots, N\}$ is a set of nodes, $E$ is a set of observed hyperedges whose elements $e \in E$ are arbitrary sets of two or more nodes in $V$, and $A$ is a vector containing the weights of edges. In this work, we assume that weights are positive and integer quantities. Denoting $\Omega$ as the set of all possible hyperedges, we have that $A_e$ is the weight of edge $e$ when $e \in E$, otherwise  $A_e = 0$ if $e\in \Omega \setminus E$. Given these definitions, the observed edge set $E$ can equivalently be represented as $E = \{e \in \Omega \, | \, A_e > 0\}$. We represent the  covariates on nodes as a matrix $X \in \mathbb{R}^{N \times Z}$, where $Z$ is the number of attributes, with entries equal to $1$ if the node $i$ has attribute $z$ and $0$ otherwise. We note that a node can have several types of covariates, e.g. gender and age, which are then one-hot encoded as attributes.

We model the presence of structural information $A$ and covariate information $X$ probabilistically, assuming a joint probability of these two types of information that is mediated by a set of latent variables $\theta=\ccup{w,\beta,u}$. Here $w,\, \beta$ are specific to each of the two distinct types of information, while the quantity $u$ is a latent variable shared between the two. The presence of a shared $u$ is a key to allow coupling the two types of information and extracting valuable insights about the system.
Formally, we assume
\begin{equation}
\label{eq: general factorization}
    P(A, X \, | \, \theta) = P_A(A \, | \, w,u) \, P_X(X \, | \, \beta,u) \,.
\end{equation}
This factorization assumes conditional independence between $A$ and $X$, given the parameters $\theta$, and is analogous to related approaches on graphs \cite{contisciani2020community,yang2013community}.  The factorization in \cref{eq: general factorization} presents various advantages. 
First, the parameters in $\theta$ can provide interpretable insights about the mechanism driving hyperedge formation, as we show below. In our case, we focus on community structure, hence we model $u$ to represent the community memberships of nodes.
Second, it allows for efficient inference of the model parameters $\theta$, as we show in the Methods section.
Third, it allows predicting both $A$ and $X$, which is relevant for example in the case of corrupted or missing data.

Having introduced the main structure of the model, we now describe the expressions of the two factors of the joint probability distribution in \cref{eq: general factorization}.

\subsection*{Modeling structural information}

We model the structural information $A$ by assuming  that latent communities control the interactions observed. For this, we utilize the \hymmsbm{} probabilistic model \cite{ruggeri2023generalized}, which assumes mixed memberships where nodes can belong to multiple communities. This model flexibly captures various community structures (e.g. assortative, core periphery etc.), scales to large hyperedge sizes and allows incorporating covariates flexibly without compromising the efficiency of its computational complexity, as we explain in the Methods section.

Assuming $K$ overlapping communities, $u$ is an $N \times K$ non-negative membership matrix, which describes the community membership for each node $i=1, \ldots, N$. A symmetric and non-negative $K \times K$ affinity matrix $w$ controls the density of hyperedges between nodes in different communities. 
The hypergraph is modeled as a product of Poisson distributions as:
\begin{equation} \label{eqn:pA}
    P_A(A | u, w) = \prod_{e\in \Omega}\,\text{Pois}\left(A_e; \frac{\lambda_e}{k_e}\right) \quad,
\end{equation}
where
\begin{equation} \label{eq2}
    \lambda_e = \sum_{i<j:i,j \in e} u_i^T w u_j = \sum_{i<j:i,j \in e} \sum_{k,q=1}^K u_{ik} u_{jq} w_{kq} \quad .
\end{equation}

The term $k_e$ is a normalization constant, which can take on any positive value. In all our experiments we set its value to
   $ k_e = \frac{|e|(|e| - 1)}{2} \binom{N - 2}{|e| - 2} $,
with $|e|$ being the size of the hyperedge. 
Other parametrizations of the likelihood $P_A(A | u, w)$ are possible, e.g. using different generative models for hypergraphs with community structure \cite{contisciani2022inference,chodrow2021generative}, but it is not guaranteed that these would yield closed-form expressions and computationally efficient algorithms when incorporating additional attribute information in the probabilistic model. Similarly, in \cref{eqn:pA} we assumed conditional independence between hyperedges given the latent variables, a standard assumption in these types of models. Such a condition could in principle be relaxed following the approaches of \cite{safdari2021generative,contisciani2022jcrep,safdari2022reciprocity}. We do not explore this here.

\subsection*{Modeling attribute information}
We model the covariates $X$ assuming that the community memberships $u$ regulate how these are assigned to nodes. We then assume that a $K \times Z$ matrix $\beta$ with entries  $\beta_{kz}$ regulates the contribution of attribute $z$ to the community $k$. This parameter plays a similar role for the matrix $X$ as the matrix $w$ does for the vector $A$.
 We combine the matrix $\beta$ with the community assignment $u$ via a matrix product that yields the following Bernoulli probabilities:
\begin{equation} \label{eq:pi}
    \pi_{iz} = \sum_{k=1}^K u_{ik}\, \beta_{kz} \quad .
\end{equation}
We assume that attributes are conditionally independent given the parameters $\pi$, which allows flexibly modeling several discrete attributes at a time. This is implemented by assuming that each entry $X_{iz}$ is extracted from a Bernoulli distribution with parameter $\pi_{iz}$ as:
\begin{equation}\label{eqn:pX}
    P_X(X| u, \beta) = \prod_{i=1}^N \prod_{z=1}^Z \pi_{iz}^{x_{iz}} (1 - \pi_{iz})^{(1 - x_{iz})} \quad.
\end{equation}
To ensure $\pi_{iz} \in [0, 1]$, we  constraint $u_{ik} \in [0, 1]$ and $\sum_{k = 1}^{K} \beta_{kz} = 1$, $\forall z$. 

We focus here on discrete and unordered attributes. This covers many relevant scenarios, including the ones we study in the several real datasets below, e.g. roles of employees in a company or classes of students. Other specific cases could be treated using similar ideas and techniques as the one we propose by suitably modifying the distribution in \cref{eqn:pX}. We give an example of imposing categorical attributes, when we want to explicitly force that having an attribute of one value does exclude any other possible value, in Supplementary Note C. 

\subsection*{Inference of latent variables}
Having defined the probabilistic model \cref{{eq: general factorization}} and the two distributions \cref{eqn:pA,eqn:pX}, our goal is to now infer the latent variables $u,w$ and $\beta$, given the observed hypergraph $A$ and the attributes $X$. To infer these values we consider maximum likelihood estimation and use an efficient expectation-maximization (EM) algorithm that
exploits the sparsity of the dataset, as detailed in the Methods section.
We combine the log-likelihoods of the two sources of information with a parameter $\gamma$ that tunes their relative contribution, with extreme values $\gamma=0$ ignoring the attributes and $\gamma=1$ ignoring the structure, similarly to what has been done in attributed network models \cite{contisciani2020community,yang2013community,fajardo2022node}, or in models for information retrieval from text \cite{hofmann1999probabilistic,zhu2013scalable}. In our experiments, we learn the $\gamma$ hyperparameter from data via cross-validation. 

Overall, the inference routine scales favorably with both the system size and the size of the hyperedges, as each EM iteration has a complexity of $O\Big( K (K + Z) (N + |E|) \Big)$, which is linear in the number of nodes and hyperedges. 
We refer to our model as \hycosbm{} and make the code available online at \repolink{}. 

\subsection*{Detecting communities in synthetic networks}

\begin{figure*}[t]
    \centering
    \includegraphics[width=1\textwidth]{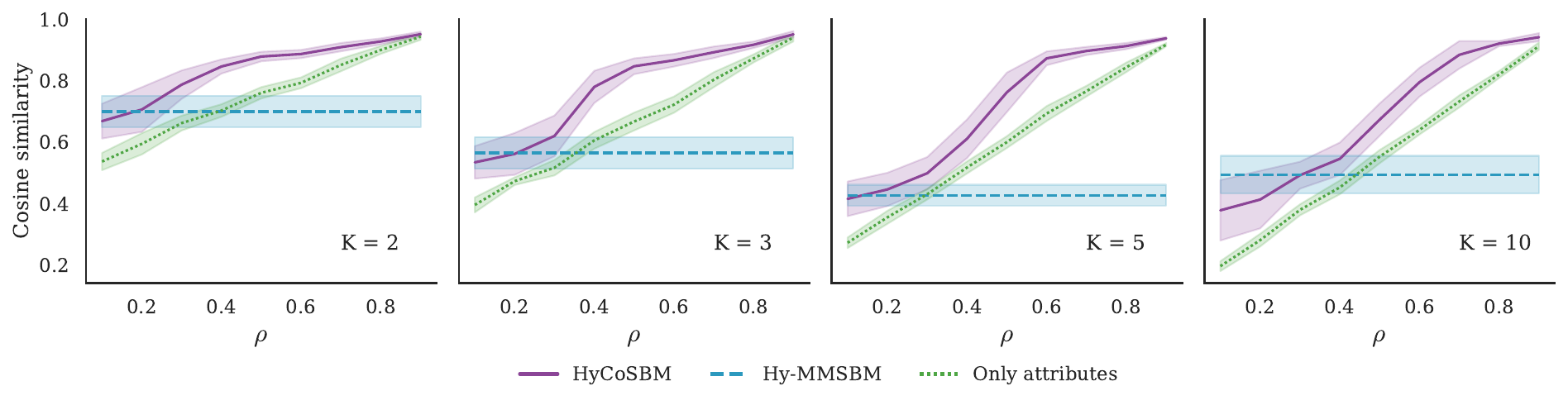}
    \caption{Community detection in synthetic hypegraphs. We show the cosine similarity between the communities inferred by the various algorithms and the ground truth communities in synthetic hypergraphs, with $N=500$ and $E = 2720$. We show results for different numbers of communities $K$ (from left to right). The number of attributes $Z$ is selected to be equal to $K$, and the parameter $\gamma$ is set equal to the fraction $\rho$ of unshuffled attributes. We compare \hycosbm{} with \hymmsbm{}, which serves as a baseline that only employs structural information. We also measure the cosine similarity of the attribute matrix $X$ and the ground truth membership matrix $u$ Only attributes). Lines and shades around them are averages and standard deviations over $10$ different network realisations.}
    \label{fig:synth}
\end{figure*} 

Our first experiments are tests on synthetic networks with known ground-truth community structure and attributes. We generate synthetic hypergraphs using \hymmsbm{}  \cite{ruggeri2023framework} as implemented in the library \texttt{HGX} \cite{lotito2023hypergraphx}. We select parameter settings where inference with \hymmsbm{} is not trivial, to better assess the influence of using attributes, see details in Supplementary Note A.
After the networks are created, we generate discrete attributes that match the community membership a fraction $\rho$ of the time, while the remaining fraction $1-\rho$ are randomly generated. This allows to vary the extent to which attributes correlate with communities and hence the difficulty of inferring the ground truth memberships. We varied $\rho\in [0.1,0.9]$, with higher values implying that  inference of communities is aided by more informative attributes.

 As a performance metric, we measure the cosine similarity between the membership vectors recovered by our model and the ground truth ones. In \cref{fig:synth} we can see that, when the attributes are correlated with ground truth communities, \hycosbm\ performs better than using either of the two types of information alone. In addition, the performance of \hycosbm\ increases monotonically with increasing correlation between attributes and ground truth. Although  this is observed also when using attributes alone, the performance of \hycosbm\ in recovering the ground truth communities is always higher.
 
 This behavior is consistent across different values of $K$, with larger performance gap between results at low and high $\rho$ at larger $K$, where there are more choices to select from.
 
 In short, these results demonstrate that the model is successfully using both attribute and structural information to improve community detection.

\subsection*{Results on empirical data}
We analyze hypergraphs derived from empirical data drawn from social, political and biological domains, as detailed in the Methods section. For each hypergraph we describe a different experiment, to illustrate various  applications of our method. We select the number of communities $K$ and the hyperparameter $\gamma$ using 5-fold cross-validation. To assess the impact of using attributes, we compare \hycosbm{} with three baselines: i) \hymmsbm, that only utilizes the structural information in the hyperedges to detect mixed-membership communities; ii) \hycosbm{} with $\gamma=0$, which is equivalent to not utilizing the attributes; iii) \hycosbm{} with community assignments $u$ fixed to match the attributes, and only infer the $w$ parameters, which tests how attributes alone perform. Notice that i) and ii) differ in that the membership vectors $u$ are unconstrained in \hymmsbm{}, while they are restricted to $u_{ik}\in [0,1]$ in our model. In iii) utilizing \hycosbm{} and \hymmsbm{} is equivalent, since the two models coincide in the updates for $w$.
The results of the following analyses are summarized in \cref{tab:results}.

Additionally, in the Supplementary Note D we
show the advantage of using a hypergraph representation by comparing against results obtained by running a probabilistic model \cite{contisciani2020community} valid on attributed pairwise networks on a clique expansion, as example dyadic representation of the datasets considered here (we refer to this approach as \clique). Notice that models valid only on pairwise data do not have a natural expression to measure the probability of a hyperedge of size larger than two. Hence one has to make an arbitrary choice on how to assign this probability from that obtained on pairwise edges. We show results for an example of this choice in the Supplementary Material. \hycosbm\
shows a strong performance in predicting hyperedges, outperforming \clique\ in all datasets except two contact datasets of students in schools, where performance is similar. Importantly,  \clique\ is limited when applied on a biological dataset with large hyperedges, as the corresponding clique expansion contains a much larger number of edges and thus creates a computational bottleneck. Overall, in the datasets considered here, we find no indication that dyadic clique
expansions are necessary neither for prediction performance nor for runtime efficiency.

\begin{table*}[hptb]
    \setlength{\tabcolsep}{4pt}
    \setlength{\arrayrulewidth}{0.05pt}
    \setlength\extrarowheight{2pt}
    \begin{tabular}{llllllcclcl}
\toprule
Dataset & Attribute & $N$ & $|E|$ & $Z$ & \multicolumn{3}{c}{\hycosbm} & \multicolumn{2}{c}{\hymmsbm} & Source \\
& & & & & $K$ & $\gamma$ & AUC & $K$ & AUC & \\
\midrule
Enron Email & structure & 4423 & 5743 & 2 & 3 & 0.700 & $0.991 \pm 0.006$ & 2 & $0.913 \pm 0.006$ & \cite{klamt2009hypergraphs} \\
\cline{1-11}
Gene Disease & DPI & 9262 & 3128 & 25 & 30 & 0.500 & $0.9 \pm 0.07$ & 2 & $0.84 \pm 0.122$ & \cite{pinero2020disgenet}\\
\cline{1-11} 
\multirow[t]{4}{*}{High School} & class & \multirow[c]{4}{*}{327}  & \multirow[c]{4}{*}{7818}  & 9 & 11 & 0.995 & $0.899 \pm 0.011$ &  \multirow[c]{4}{*}{24} & \multirow[c]{4}{*}{$0.884 \pm 0.006$} &  \multirow[t]{4}{*}{\cite{genois2018can}}\\
 & has filled questionnaire & &  & 2 & 21 & 0.800 & $0.892 \pm 0.013$ &  &  \\
 & has facebook &  &  & 2 & 15 & 0.950 & $0.888 \pm 0.008$ &  &  \\
 & sex &  &  & 2 & 16 & 0.800 & $0.889 \pm 0.009$ &  & 
 \\
\cline{1-11}
\multirow[t]{2}{*}{Primary School} & class & \multirow[c]{2}{*}{242} &  \multirow[c]{2}{*}{12704}  & 11 & 10 & 0.600 & $0.841 \pm 0.013$ & \multirow[c]{2}{*}{11} & \multirow[c]{2}{*}{$0.841 \pm 0.007$} \\
 & sex &  &  & 2 & 12 & 0.200 & $0.841 \pm 0.007$ & & & \multirow[t]{2}{*}{\cite{genois2018can}}\\
\cline{1-11}
Hospital & status & 75 & 1825 & 4 & 2 & 0.200 & $0.776 \pm 0.032$ & 2 & $0.758 \pm 0.016$& \cite{genois2018can} \\
\cline{1-11}
Workplace & department & 92 & 788 & 5 & 5 & 0.995 & $0.81 \pm 0.02$ & 5 & $0.752 \pm 0.039$ &  \cite{genois2018can}  \\
\cline{1-11}
House Bills & political party & 1494 & 54933 & 2 & 22 & 0.000 & $0.952 \pm 0.003$ & 25 & $0.952 \pm 0.001$ &\cite{fowler2006connecting, fowler2006legislative}\\
\cline{1-11}
House Committees & political party & 1290 & 335 & 2 & 13 & 0.100 & $0.985 \pm 0.015$ & 24 & $0.972 \pm 0.011$&\cite{stewart2008congressional} \\
\cline{1-11}
Senate Bills & political party & 294 & 21721 & 2 & 23 & 0.000 & $0.929 \pm 0.006$ & 19 & $0.923 \pm 0.003$ &\cite{fowler2006connecting, fowler2006legislative}\\
\cline{1-11}
Senate Committes & political party & 282 & 301 & 2 & 23 & 0.000 & $0.972 \pm 0.01$ & 21 & $0.963 \pm 0.023$&\cite{stewart2008congressional} \\
\bottomrule
\end{tabular}
    \caption{AUC scores on real datasets. We report the AUC scores resulting from 5-fold cross-validation on various real datasets. Values and errors are averages and standard deviations over 5 cross-validation folds. We report the number of nodes $N$, number of hyperedges $|E|$, number of attributes $Z$ and  the values of $K$ and $\gamma$ as obtained from cross-validation. }
    \label{tab:results}
\end{table*}


\subsection*{Recovering interactions on contact dataset}
In our first experiment we study human contact interactions, using the data obtained from wearable sensor devices in four settings \cite{mastrandrea2015contact,gemmetto2014mitigation, stehle2011high,genois2015data,vanhems2013estimating}: students in a high school (High School) and a primary school (Primary School), co-workers in a workplace (Workplace) and patients and staff in a hospital (Hospital). Hyperedges represent a group of people that were in close proximity at some point in time. Each dataset contains attributes that describe either the classes, the departments, or the roles the nodes belong to. 

We measure the ability of our model to explain group interactions by assessing its performance on a hyperedge prediction task. To this end, we infer the parameters using only a fraction of the hyperedges in the dataset. Then we utilize the held out hyperedges to measure the AUC metric, which represents the fraction of times the model predicts an observed interaction as more likely than a non-observed one (higher values mean better performance).

Models that do not utilize any attribute have been previously shown to perform well on such a task on these datasets \cite{chodrow2021generative,contisciani2022inference,ruggeri2023generalized} when a large fraction of the dataset was given as input. Here, we vary the amount of structural information available to the algorithms more pronouncedly to assess their robustness in realistic situations where the full data is unavailable and investigate how making use of attributes can compensate for this. To simulate this setting, we delete an increasing fraction of the existing hyperedges (keeping the hypergraph connected) and perform 5-fold cross-validation on the remaining dataset.

\begin{figure}
    \centering
    \includegraphics[width=0.50\textwidth]{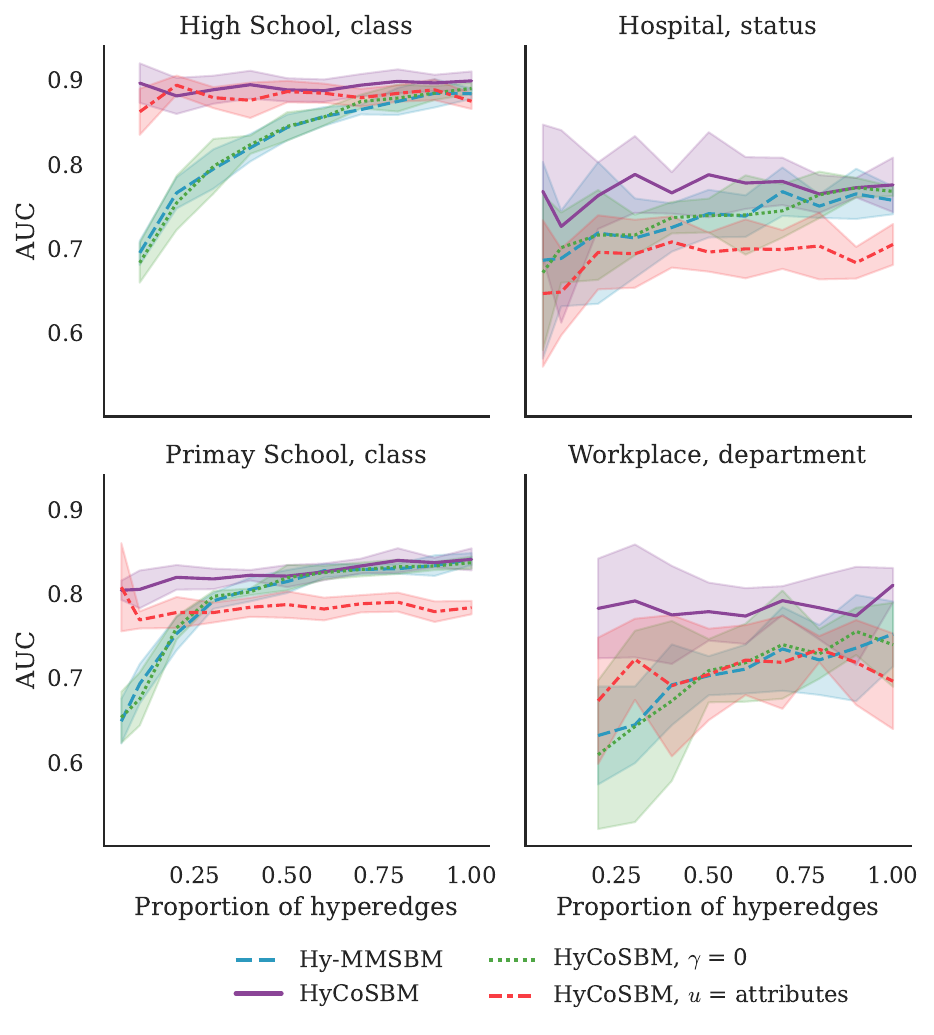}
    \caption{Predicting interactions in close-proximity datasets with partial observations. We show the performance of various methods in hyperedge prediction tasks, measured by AUC, as we vary the fraction of hyperedges made available to the algorithms. This plot shows that the performance of \hycosbm{} remains high when fewer hyperedges are available in input, while that of the algorithms which do not use any attribute drops. Lines and shades around them are averages and standard deviations over $5$ cross-validation folds.}
    \label{fig:contacts-partition}
\end{figure}

The results in \cref{fig:contacts-partition} show a significant and monotonic drop in performance for 
\hymmsbm{} as we decrease the fraction of hyperedges, consequently reducing the amount of structural information available to the algorithm.
In contrast, \hycosbm{} maintains an almost constant and high performance, all the way down to having access only to $20\%$ of the hyperedges, owing to its usage of the additional attribute information. In addition, even in the favorable setting when all hyperedges are available, \hycosbm{} yields higher AUC in Workplace (with $\gamma = 0.995$), indicating that incorporating attributes can be beneficial even when robust results are obtained using structural information alone.

Focusing on other datasets where \hycosbm{} attains AUC similar to that of other algorithms when all the interactions are utilized, we still observe a difference in the types of communities detected. As an example, in the High School dataset the community assignments $u$ inferred via \hymmsbm{} have cosine similarity of $0.59$ with the class attribute of the nodes, as opposed to the cosine similarity of $0.94$ observed for \hycosbm{}.

These different levels of correlation between inferred communities and attributes, together with observing similar AUC (indicating a similar ability to explain the structural information), could be explained by the presence of competing network divisions, as already observed in network datasets \cite{good2010performance, newman2016structure,peel2017ground}.  
Our model allows selecting among divisions, finding ones that correlate with the attribute of interest.  

\begin{figure*}[t]
    \centering
    \includegraphics[width=1\textwidth]{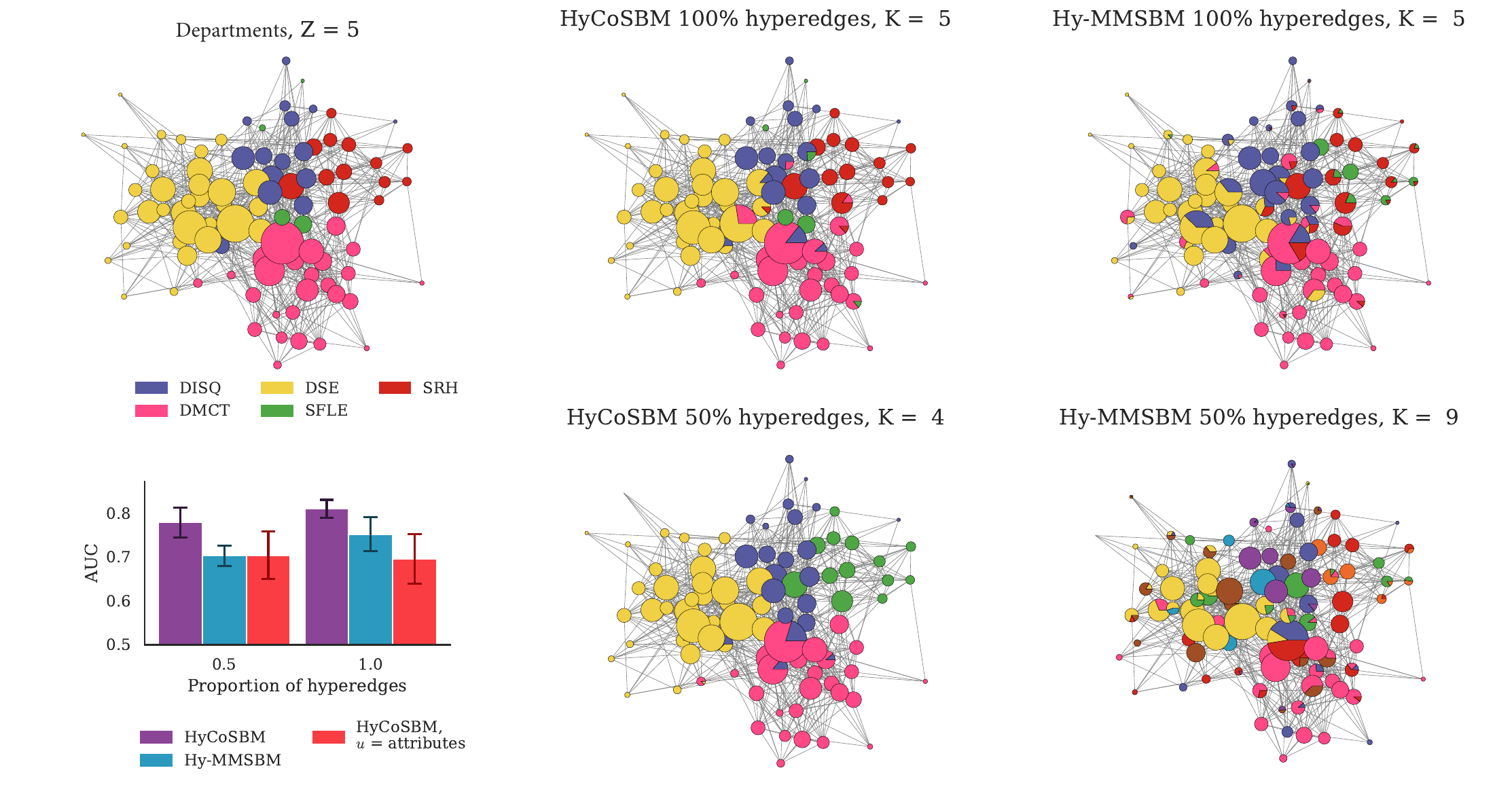}
    \caption{Communities detected in a Workplace dataset from partial observations of close-proximity interactions. We vary the fraction of hyperedges given in input to the algorithms (top: $100\%$, bottom: $50\%$) and compare the inferred communities against the attribute departement (top left). The AUC barplot (bottom-left) shows the performance of the models in hyperedge prediction. Bars and error bars are averages and standard deviations over $5$ cross-validation folds. This plot shows that \hycosbm\ is able to use the attributes effectively to keep performance high even at a low fraction of input observations. }
    \label{fig:workplace}
\end{figure*} 

We highlight that, although the communities inferred by \hycosbm{} correlate with the attributes, these two are not equivalent. In fact, we observe several cases where the number of detected communities is not equal to the number of  attributes. 
For example, we observe cases where the model detects fewer communities than the number of attributes available. In \cref{fig:workplace}  the nodes with attribute SFLE (green) are included within the  community formed mainly by DISQ nodes (purple) by our model when $50\%$ of the edges are given in input. This partition achieves higher AUC than the model with community assignments fixed and equal to the attributes. In other cases, our model finds smaller communities within the bigger partitions determined by the attributes. We find such an example in the High School dataset in Supplementary Figure 1, where \hycosbm{} finds finer partitions ($K=11$) than the one given by the $Z=9$ classes, hierarchically splitting some classes into subgroups. 
The resulting partition attains a high AUC score. A high number of inferred communities is also observed in \hymmsbm{}, but, in this case, the AUC drops significantly, and the $K=30$ communities inferred at $30\%$ of the edges are much more mixed between the classes.
In short, the communities inferred by our model do not simply replicate the attribute. Rather, this additional information is used to infer a community structure that better explains the interactions observed in the data.

\begin{figure}[t]
    \centering
    \includegraphics[width=0.50\textwidth]{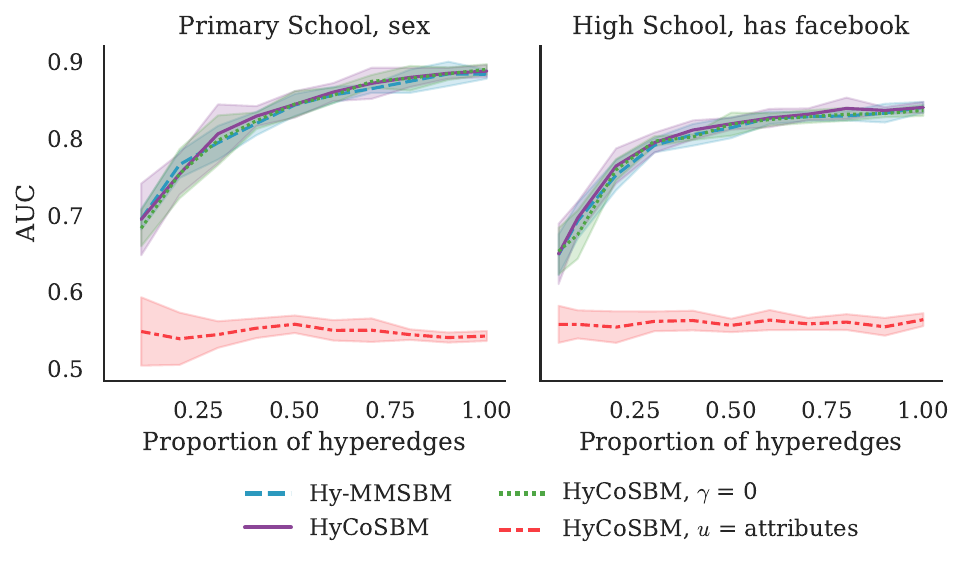}
    \caption{AUC on contacts dataset with partial hyperedges: uncorrelated attributes. Using \texttt{sex} and \texttt{has facebook} as the attributes, the performance of all models drops as the hyperedges are removed. Lines and shades around them are averages and standard deviations over $5$ cross-validation folds.}
    \label{fig:bad-attributes}
\end{figure}

\subsection*{Performance with uninformative attributes}
In the previous sections, we have shown how attribute information can aid the recovery of effective communities and improve inference.
In general, though, we cannot expect that any type of attribute added to a network dataset may help explaining the observed structure. This may be the case for instance when an attribute is uncorrelated or weakly correlated with the hyperedges, as in the synthetic experiments described above when $\rho$ is close to $0.1$. 

In this section we study the performance of \hycosbm{} in this adversarial regime and show that, when attributes are uninformative, these are readily discarded by our model to only perform inference based on structural information.

To this end, we feed the \texttt{sex} and \texttt{has facebook} attributes, respectively from the Primary School and High School datasets, into our model. 
As we show in \cref{fig:bad-attributes}, the performance of \hycosbm{} closely resembles that of the models that do not use any attribute in input, signaling that these attributes are not as informative as \texttt{class} to explain the observed group interactions. This is reinforced by a very low AUC for the model that fixes $u$ as the attributes (red line). 

We further illustrate this point in four datasets of US representatives. Here, nodes are representatives (in the House of Representatives or in the Senate) and hyperedges represent co-sponsorship of bills (Bills datasets) or co-participation in a committee during a Congress meeting (Committees datasets).
The attribute indicates whether the representative is associated with the Republican or Democratic party ($Z=2$). In \cref{tab:UScovoting} we show that there is no advantage in using this binary attribute to explain the co-sponsorship nor the co-participation patterns, as the AUC is similar to that of models that do not use attribute information in input. As a confirmation, the value of $\gamma$ obtained via cross-validation is equal to $0$ in three out of four cases, and $0.1$ in one case, showing that the algorithm tends to discard the attribute information and prefers to rely solely on structural data. 

\begin{table}[hptb]
\small
    \centering
    \setlength{\tabcolsep}{3pt}
    \setlength{\arrayrulewidth}{0.1pt}
    \begin{tabular}{lrrcrc}
\toprule
Dataset & \multicolumn{3}{c}{\hycosbm} & \multicolumn{2}{c}{\hymmsbm} \\
&   $K$ & $\gamma$ & AUC & $K$ & AUC \\
\midrule
House Bills & 22 & 0.0 & $0.952 \pm 0.003$ & 25 & $0.952 \pm 0.001$ \\
House Committees &  13 & 0.1 & $0.985 \pm 0.015$ & 24 & $0.972 \pm 0.011$ \\
Senate Bills &  23 & 0.0 & $0.929 \pm 0.006$ & 19 & $0.923 \pm 0.003$ \\
Senate Committees  & 23 & 0.0 & $0.972 \pm 0.01$ & 21 & $0.963 \pm 0.023$ \\
\bottomrule
\end{tabular}
    \caption{AUC scores on co-sponsorship and co-participation datasets of US representatives. We report the results of cross-validation in terms of selected $K$, $\gamma$, and obtained AUC. Here the node attribute used by \hycosbm{} is the political party of the representative (Democrat or Republican, $Z=2$).}
    \label{tab:UScovoting}
\end{table}

\subsection*{Improving prediction of Gene-Disease associations}

\begin{figure}[h]
    \centering
    \includegraphics[width=0.50\textwidth]{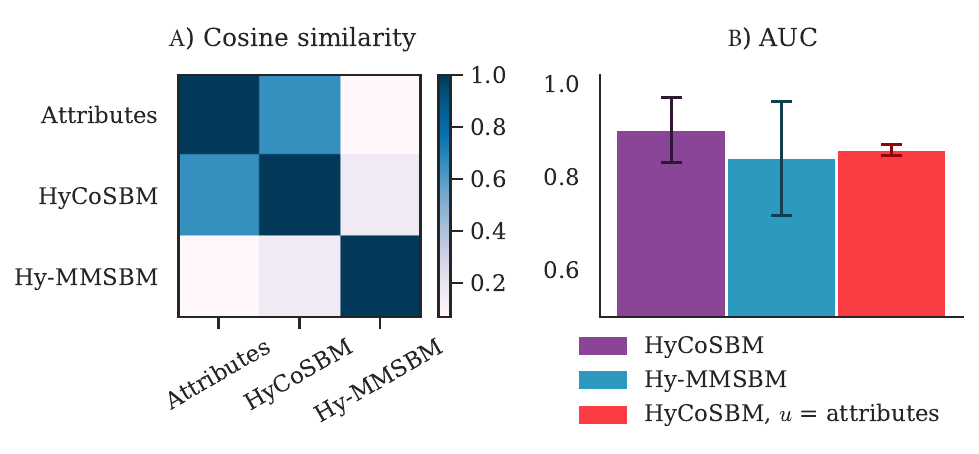} 
    \caption{Cosine similarity and AUC in a Gene Disease dataset. A) Cosine similarity between the three types of communities: attribute, \hycosbm\ and \hymmsbm. B) AUC in predicting missing hyperedges. Bars and error bars are averages and standard deviations over $5$ cross-validation folds. The membership $u$ detected by \hycosbm\ correlates with the DPI attribute and achieves higher AUC than both \hymmsbm\  and the model trained with $u$ fixed as the attribute.
    }\label{fig:gene-desease-cossim}
\end{figure}

Our next application is on a biological dataset containing Gene-Disease associations \cite{pinero2020disgenet}. 
Here, nodes represent genes, and hyperedges represent a combination of genes specific to a disease. For each node, its Disease Pleiotropy Index (DPI) is available as an attribute, indicating the tendency of a gene to be associated with many types of diseases, with $Z=25$ possible discrete values.
The dataset is highly sparse, as many nodes are present only in one hyperedge. Previous results have shown that inferring missing associations improves sensibly when using all hyperedges in the datasets \cite{ruggeri2023generalized} (with AUC scores up to $0.84$), compared to using  only hyperedges up to size $D=25$ \cite{contisciani2022inference}. In this paragraph, we investigate whether these results can be further improved when additional information is available in the form of the DPI attribute. 
We find that running \hycosbm{} achieves an AUC score of $0.9$, indicating that this attribute is informative. Furthermore, we observe that the communities detected by \hycosbm{} are similar to those obtained from the attributes, see \cref{fig:gene-desease-cossim}a), but with a finer division into $K=30$ communities, which is larger than the $Z=25$ covariate categories. 

\subsection*{Recovering core-periphery structure with Enron Email dataset}
In this paragraph we focus on the application of our methodology to the Enron Email dataset \cite{klimt2004enron}, where nodes represent employees of an organization and hyperedges email exchanges. In particular, the dataset comes with the annotation of nodes being either part of a ``core", which contains employees sending batch emails, or a ``periphery", containing the receivers. A hyperedge represents one email batch, and it contains both the core sender and all the periphery receivers. Here, we focus on the study of the core-periphery attribute to predict higher-order interactions in the data.

In this dataset, the core-periphery generative process behind the data is partially known. Hence, it does not come as a surprise that using the ``core'' and ``periphery'' labels improves inference and reconstruction. However, our results using \hycosbm\ reveal both a more effective and a more nuanced interpretation than that given by using the labels alone. This is because \hycosbm\ does not simply replicate the attributes of the nodes, but rather exploits them to achieve an improved inference. To test this hypothesis, we compare three inference scenarios: the vanilla \hycosbm\ inference, a constrained version where the attribute matrix $u$ is fixed and equal to the core-periphery assignments, and the \hymmsbm\ algorithm. These achieve AUC scores of $0.99,  0.95$ and $0.91$, respectively.

\begin{figure*}[t]
	\centering
	\includegraphics[width=1\textwidth]{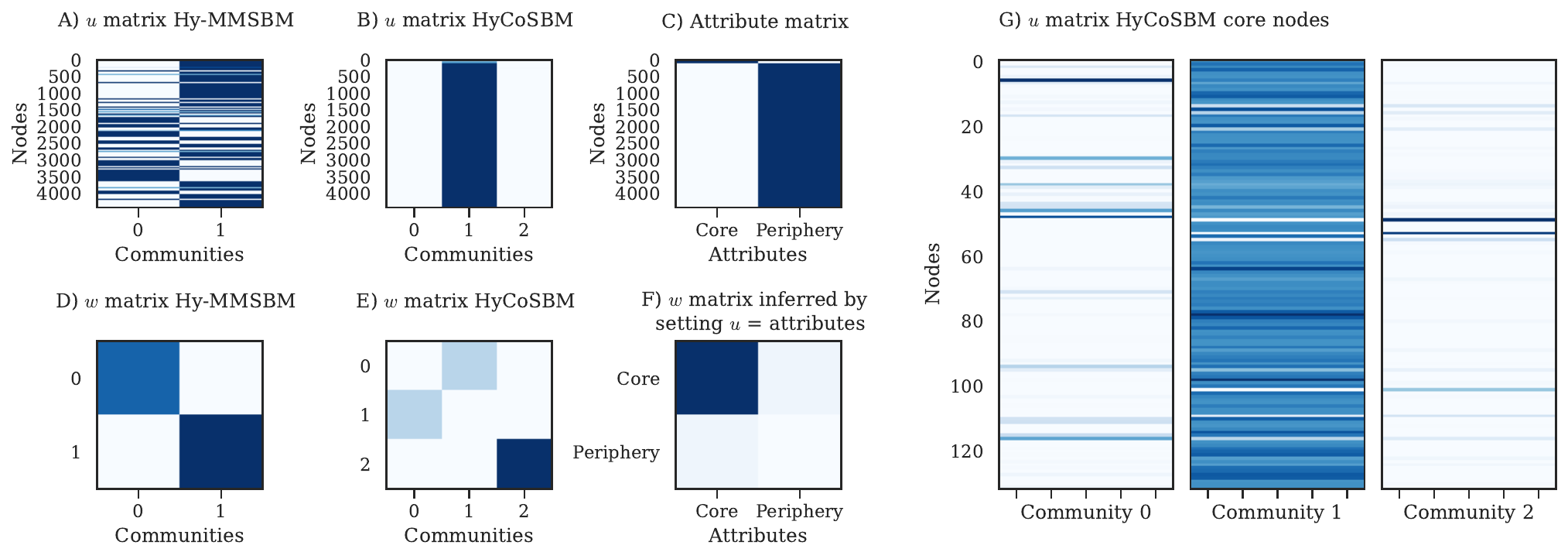}
	\caption{Communities and structure detected in the Enron-Email dataset.(A,B,C): the $u$ matrices inferred by \hymmsbm,  \hycosbm\  and the attributes matrix. The rows $u_{i}$ for \hymmsbm\ are shown normalized as $u_{i}/\sum_{k}u_{ik}$ for better clarity.  (D,E,F): the $w$ matrices inferred by \hymmsbm, \hycosbm\ and by  \hycosbm\ when fixing $u$ to the attributes matrix. G) Zooming in the $u$ matrix for \hycosbm\ to highlight the mixed-membership of core nodes. We notice how core nodes have mixed-membership spanning two or three groups. Periphery nodes instead mostly belong only to Community 1 (not shown here).  }
	\label{fig:enron}
\end{figure*}

As it can be observed in \cref{fig:enron}, \hymmsbm{} finds assortative structure dividing the nodes into 2 groups, which is also the number of attributes. Instead, \hycosbm\ divides the nodes into three groups: groups 0 and 1, which interact with each other in a disassortative fashion, and group 2, that behaves assortatively. We also observe that the large majority of nodes that have mixed-membership spread in these three groups are core nodes, while periphery nodes have mainly a non-zero membership in group 1. As a result,  \hycosbm\  unveils a finer-grain division of the core, revealing patterns within it that cannot be inferred by observing the (hard) membership given by the attributes themselves.  This is also shown by the inferred $w$ matrix, where core nodes interact mainly with themselves and partially with periphery nodes when we fix $u$ equal to the labels.

In summary, \hycosbm\ effectively leverages the data attributes to inform the inference procedure. It does so by exploiting the additional information to extract informative structure and unveiling finer structure than the one given by the observed attributes alone.

\subsection*{Predicting co-destination patterns in New York City taxi rides}
As a final application, we consider a dataset of taxi rides in New York City \cite{nycdata}. We are interested in measuring patterns of similar destinations, based on travel demands. For this, we consider a given time window and a day of the week and build a hypergraph where nodes are dropoff locations and a hyperedge connects dropoffs that where reached by travellers starting from the same pickup location. Data of this form is often used in urban planning and to understand human mobility co-location patterns \cite{yang2015poi,chen2022contrasting}. The only node attribute available from the data is the ``Borough'' type (the basic administrative unit in the city of New York), which we utilize in the following experiments. In addition to the existing five boroughs, the dataset also contains Newark airport as location. We assign it to a $6$th attribute.

We study examples of such a network by considering the week from Saturday 04-11-2023 until Tuesday 14-11-2023, and building two hypergraphs relative to two different time windows: i) Monday and Tuesday 06-07.11 between 17.00 and 20.00, and ii) Saturday and Sunday 04-05.11 between 00.00 and 03.00. These two $3$-hour time windows are selected to consider diverse travel needs. We expect the first one to capture commuters, the second to capture entertainment and nightlife. We obtain two hypergraphs with $N=214$ nodes, $E=523$ and $476$ hyperedges, and maximum hyperedge sizes of $132$ and $125$, respectively, see \cref{tab:nyc_stats}.\\

\begin{table}[htbp]
\centering
\setlength{\tabcolsep}{3pt}
\setlength{\arrayrulewidth}{0.1pt}
\begin{tabular}{lrrrr}
	\toprule
	Dataset & $N$ & $|E|$ & $|E_{2}|$ & $|e|_{max}$ \\
	\midrule
	Mon-Tue 17-20  & 214 & 523 & 64 & 132 \\
	Sat-Sun 00-03  & 214 & 476 & 53 & 125\\
	\bottomrule
\end{tabular}
    \caption{Statistics on hypergraph obtained from NYC taxi drives. Number of nodes $N$, number of hyperedges $|E|$,  number of dyadic hyperedges $|E_{2}|$ and maximum hyperedge size $|e|_{max}$.}
\label{tab:nyc_stats}
\end{table}

We assess how informative \hycosbm\ is in representing co-destination taxi trips data by comparing it with other approaches in the task of predicting future co-destination locations. Specifically, we train a model on the two datasets described above, and perform hyperedge prediction on analogous datasets built from taxi rides taking place in subsequent days of the same week, in two different time windows. While we expect travel demands to vary with time and day, we also expect correlations to be exploited because of the intrinsic nature of different destination locations, which could attract similar types of passengers in different times and days. To better test this hypothesis, we devise an experiment where for each existing hyperedge $e$ in a given test set (e.g. for the taxi trips of Wednesday and Thursday between 00.00 and 00.03), we extract a non-existing hyperedge $\hat{e}$ where we make a minimal change to $e$. Specifically, we select one node $i \in e$ at random and switch it with another node $j \in V \setminus {e}$, also selected at random. We refer to this procedure as switch-one-out (SOO). In this way we make the task more difficult as all nodes but one coincide in $e$ and $\hat{e}$. In terms of the Jaccard similarity, we have $J(e,\hat{e}) = |e \cap \hat{e}|/|e \cup \hat{e}|=|e-1|/|e+1|$.
This construction of the negative test data aims at building challenging comparisons as the prediction of true positives becomes more difficult due to $e$ and $\hat{e}$ being similar.
We first run cross-validation on the training datasets and choose the best parameters. Then we analyze the results using test datasets generated from a different day.

\begin{figure*}[t]
	\centering
	\includegraphics[width=1\textwidth]{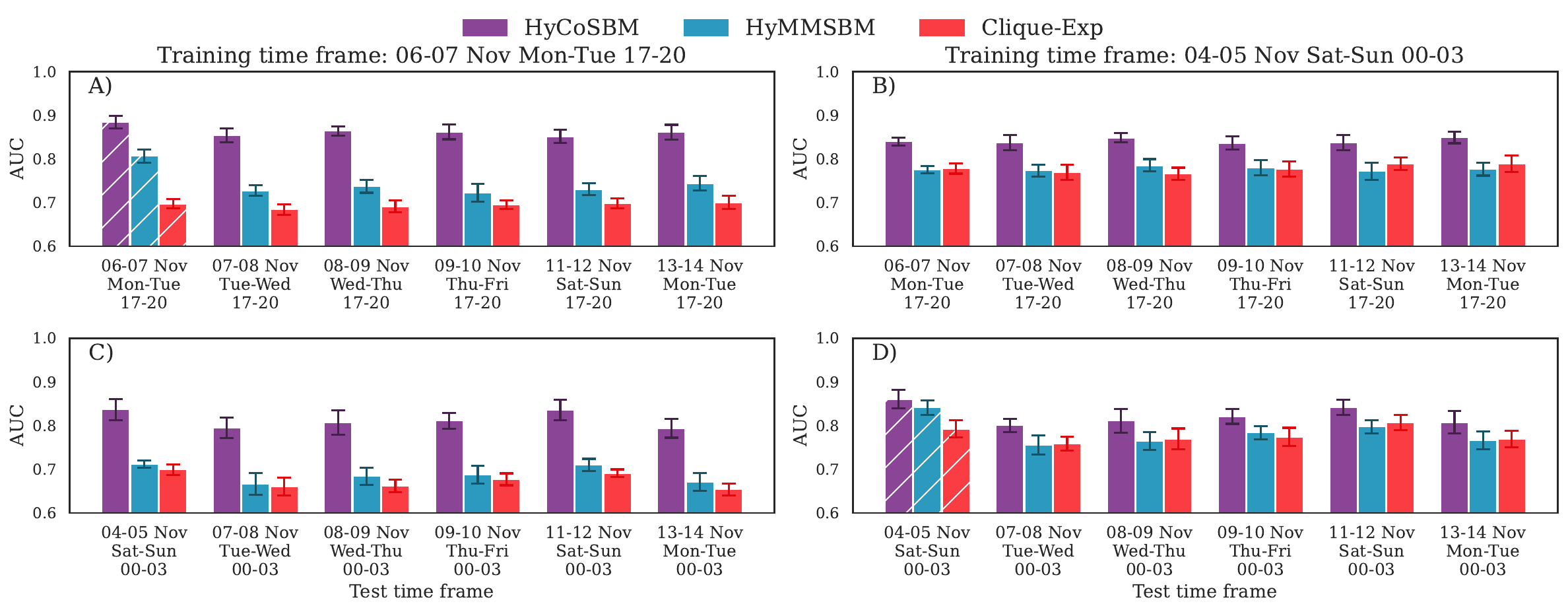}
	\caption{Predicting co-destination taxi rides in New York City.  We report the AUC calculated running SOO hyperedge prediction on hypergraph test sets built considering taxi rides taking place in various time frames subsequent to the two used to train the algorithm. Test time frames with time window between 17.00-20.00 (A-B) and 00.00-03.00 (C-D), for the two training sets (left-right). Hatched bars denote the performance in the training sets; bars and errors are average and standard deviations over 10 random realisations of the SOO procedure.}
	\label{fig:nyc_auc}
\end{figure*}

Observing \cref{fig:nyc_auc}, we find that \hycosbm\ achieves a strong performance in predicting future co-destinations consistently across test time frames, which is significantly higher than that of the two comparison methods, \hymmsbm\ and \clique. The performance gap is higher in predicting co-destinations using the time window of Monday and Tuesday between 17.00 and 20.00 as training set, where the other two approaches have much lower AUC, with \hymmsbm\ attaining higher values than \clique. We find that \hycosbm\ detects communities that are partially aligned with the ``Borough'' attribute (not shown here).
Furthermore, we observe that various node are assigned with mixed-membership spread over more than one community, and that several communities comprise nodes from different boroughs.

\section*{Discussion}
\label{sec: discussion}
We have analyzed how node attributes can be used to guide investigations of higher-order data. We focused on the problem of community detection, introducing a mixed-membership probabilistic generative model for hypergraphs. Our model can explicitly incorporate both hyperedges and node attributes, and find more expressive community partitions by exploiting the combination of these information sources.

We have applied our model to a variety of social, political and biological hypergraphs, showing how prediction of missing interactions can be boosted by the addition of informative attributes, in particular in the regime of incomplete or noisy data. We have also illustrated various scenarios where attributes can be used to select between competing divisions, or cases where they are not informative and can be discarded.

There are a number of possible extensions of this work. One could include additional attribute types, such as attributes on hyperedges, continuous variables or vector variables, for instance considering recent approaches for attributed networks \cite{contisciani2024flexible}. Similarly, one could consider alternative probabilistic expressions for the structural data, but this would require efforts to derive closed form updates and maintain a low computational complexity. On a related note, our model is based on the assumption that attributes and structure are independent conditionally on the latent variables. This approach is rather general, as the latent variables can potentially take on different semantics. It would be interesting to study other types of dependencies between structure and attributes, as well as investigating in more depth the validity of conditional dependence assumptions in both hyperedges and attributes.
Finally, our model might be extended to consider dynamical hypergraphs, where communities and interactions can change in time, and assess what role attributes play in this case. 

\section*{Methods}
\subsection*{Inference of the latent variables}
The likelihood of \hycosbm{} factorizes over all hyperedges $e \in \Omega$, and single hyperedges are modeled with a Poisson distribution:
\begin{equation} \label{eqnM:pA}
    P_A(A_e | u, w) = \text{Pois}\left(A_e; \frac{\lambda_e}{k_e}\right) \, .
\end{equation}
Similarly, the probability of attributes factorizes into Bernoulli probabilities:
\begin{equation}\label{eqnM:pX}
    P_X(X| u, \beta) = \prod_{i=1}^N \prod_{z=1}^Z \pi_{iz}^{x_{iz}} (1 - \pi_{iz})^{(1 - x_{iz})} \quad.
\end{equation}
Under the Poisson distribution in \cref{eqnM:pA}, it can be shown that the log-likelihood $L_A(u, w)$ of the full hypergraph evaluates to 
\begin{equation}\label{eq: loglik structure}
    L_A(u, w) = -C \sum_{i<j \in V} u_i^T w u_j + \sum_{e \in E} A_e \log \sum_{i<j \in e} u_i^T w u_j\, ,
\end{equation}
where $C = \sum_{d=2}^D \binom{N-2}{d-2} \frac{1}{\kappa_d}$ and $D$ is the maximum hyperedge size observed \cite{ruggeri2023generalized}.
Instead,   \cref{eqnM:pX}
yields the log-likelihood
\begin{align}
    \notag L_X(u, \beta) & = \sum_{i=1}^N \sum_{z=1}^Z x_{iz} \log \left( \sum_{k=1}^K u_{ik}\, \beta_{kz} \right) \\
    & + \sum_{i=1}^N \sum_{z=1}^Z (1 - x_{iz}) \log \left( \sum_{k=1}^K (1 - u_{ik}) \,\beta_{kz} \right) \quad .
\end{align}

As we assumed conditional independence of the network part and the attributes part, the total log-likelihood becomes the sum of those two terms. In practice though, performance improves by introducing a balancing parameter $\gamma \in[0,1]$ that tunes the relative contribution of the two terms \cite{contisciani2020community,fajardo2022node,zhu2013scalable, hofmann1999probabilistic},
yielding a total log-likelihood as:
\begin{align} \label{eq:total_ll}
    L(u, w, \beta) = (1 - \gamma)\, L_A (u, w) + \gamma \, L_X (u, \beta) \quad.
\end{align}
The value of $\gamma$ is not known a priori, and it can be learned from the data using standard techniques for hyperparameter learning. In our experiments, we utilize cross-validation. 
The $\gamma$ parameter is necessary to better balance the contribution of the structural and covariate information, as the magnitude of the two different log-likelihood terms can be on different scales, with the risk of biasing the total likelihood maximization towards one of the two terms. This balancing is also useful when attribute data are somehow more (or less) reliable than structural data, for instance when we believe that one is less (or more) subject to noise. Furthermore, $\gamma$ is reminiscent of any hyperparameter of approaches that adjust inference based on prior distributions on the community assignments, as done in some attributed network models, e.g. \cite{newman2016structure,tallberg2004bayesian}.

We note here that the value of $\gamma$ has a clear interpretation only for the extreme cases of $0$ or $1$, which discards entirely the contribution of one of the two terms. In all the other intermediate cases,  its value is not simply interpreted as a percentage contribution of the attributes over the network. This is because $\gamma$  balances the magnitudes of two likelihood terms. In general, the network part is much larger than the attribute one, which draws $\gamma$ to values closer to $1$, e.g. $0.995$, to compensate for  the difference in scales. This does not necessarily mean that the network information is barely used, but rather that it has to be rescaled to allow the attribute information to be effectively considered.

As a final remark, our definition of $X$ allows modeling several discrete attributes at the same time, and the dimension $Z$ is the total number of values, including all the attribute types. Formally, $Z=\sum_{p=1,\dots,P}z_p$, where $P$ is the number of attribute types (e.g. age and class would give $P=2$), and $z_p$ is the number of discrete values an attribute of type $p$ can take. Alternatively, the presence of more than one attribute can be modeled by considering separate terms $L_X$, each with a different multiplier $\gamma$. While this formulation would allow for tuning the contribution of attributes more specifically, this comes at a price of higher model complexity (in case of using different expressions for the $L_X$) or higher computational complexity, as one needs to cross-validate more than one type of $\gamma$. We do not explore this here.

\subsubsection*{Variational lower bound}
To maximize the total log-likelihood in \cref{eq:total_ll} we adopt a standard variational approach to lower bound the summation terms inside the logarithm. Introducing the probability distributions $\rho_{ijkl}^{(e)}, h_{izk}$ and $h'_{izk}$ and using Jensen's inequality $\log \mathbb{E}[x] \geq \mathbb{E} [\log x]$, we get the following lower bounds:
\begin{align}
    \notag &\sum_{e \in E} A_e \sum_{i<j \in e} \log \sum_{k,q=1}^K \left(u_{ik} u_{jq} w_{kq} \right) \geq  \\
    &\quad \sum_{e \in E} A_e \sum_{i<j \in e} \sum_{k,q=1}^K \rho_{ijkq}^{(e)} \log \left(\frac{u_{ik} u_{jq} w_{kq}}{\rho_{ijkq}^{(e)}}\right) \,;  \label{eq: variational ineq 1} \\
    \notag &\sum_{i=1}^N \sum_{z=1}^Z x_{iz} \log \left( \sum_{k=1}^K u_{ik} \beta_{kz} \right) \geq \\
     & \quad \sum_{i=1}^N \sum_{z=1}^Z x_{iz} \sum_{k=1}^K h_{izk} \log \left(  \frac{u_{ik} \beta_{kz}}{h_{izk}} \right)\,; \label{eq: variational ineq 2}\\
    \notag &\sum_{i=1}^N \sum_{z=1}^Z (1 - x_{iz}) \log \left( \sum_{k=1}^K (1 - u_{ik}) \beta_{kz} \right) \geq \\
    &\quad \sum_{i=1}^N \sum_{z=1}^Z (1 - x_{iz}) \sum_{k=1}^K h'_{izk} \log \left(  \frac{(1 - u_{ik}) \beta_{kz}}{h'_{izk}} \right) \,; \label{eq: variational ineq 3}
\end{align}
with equality reached when
\begin{align}
    \rho_{ijkq}^{(e)} &= \frac{u_{ik} u_{jq} w_{kq}}{\lambda_e}\,; \label{eqn:rho}\\
    h_{izk} &= \frac{\beta_{kz} u_{ik}}{\sum_{k'} \beta_{k'z} u_{ik'}} \label{eqn:h}\,;  \\
    h'_{izk} &= \frac{\beta_{kz} (1 - u_{ik})}{\sum_{k'} \beta_{k'z} (1 - u_{ik'})}\, \label{eqn:h1};
\end{align}
respectively.

Plugging \cref{eq: variational ineq 1} into \cref{eq: loglik structure} yields a lower bound $\mathcal{L}_A$ of the structural log-likelihood
\begin{align}
    &\notag \mathcal{L}_A(u, w, \rho) = -C \sum_{i<j \in e} u_i^T w u_j  \\ 
    & \hspace{0mm} + \sum_{e \in E} A_e \sum_{i<j \in e} \sum_{k,q=1}^K \rho_{ijkq}^{(e)} \log \left(\frac{u_{ik} u_{jq} w_{kq}}{\rho_{ijkq}^{(e)}}\right) \, .
\end{align}
Similarly, \crefrange{eq: variational ineq 2}{eq: variational ineq 3} yield a lower bound $\mathcal{L}_X$ of the log-likelihood of the attributes:
\begin{align}
    & \mathcal{L}_X(u, \beta, h, h') = \sum_{i=1}^N \sum_{z=1}^Z x_{iz} \sum_{k=1}^K h_{izk} \log \left(  \frac{u_{ik} \beta_{kz}}{h_{izk}} \right)  \nonumber \\
    & \hspace{0mm} + \sum_{i=1}^N \sum_{z=1}^Z (1 - x_{iz}) \sum_{k=1}^K h'_{izk} \log \left(  \frac{(1 - u_{ik}) \beta_{kz}}{h'_{izk}} \right) \, ,
\end{align}
so that 
\begin{equation}\label{eq: full variational lower bound}
    \mathcal{L} := (1 - \gamma) \mathcal{L}_A + \gamma \mathcal{L}_X \, ,
\end{equation}
is a lower bound of the full log-likelihood.

\subsubsection*{Expectation-Maximization}
We now aim to optimize the variational lower bound in \cref{eq: full variational lower bound} with respect to the model parameters $u, w$ and $\beta$. To account for the constraint on $\beta$ and $u$, we introduce the Lagrange multipliers $\lambda^{(\beta)}$ and $\lambda^{(u)}$ obtaining the following objective:
\begin{align}
    \mathcal{L}_{constr} := \mathcal{L}
    - \sum_{z=1}^Z \lambda_z^{(\beta)} \left ( \sum_{k=1}^K \beta_{kz} - 1 \right ) - \sum_{i}^N \sum_{k}^K \lambda_{ik}^{(u)} u_{ik} \, . \label{eq: objective with lagrange}
\end{align}

We proceed as in the Expectation-Maximization algorithm \cite{dempster1977maximum}, by alternating two optimization steps until convergence. In one step, we maximize \cref{eq: objective with lagrange} with respect to the model parameters $u, w, \beta$ and the Lagrange multipliers $\lambda^{(\beta)}, \lambda^{(u)}$. In the other, we utilize the closed-form updates in \crefrange{eqn:rho}{eqn:h1} for the variational parameters. The procedure is described in detail in \cref{pseudocode:EM}.

Differentiating objective \cref{eq: objective with lagrange} with respect to the $w, \beta$ parameters and the multipliers $\lambda^{(\beta)}$ yields the following closed-form updates:
\begin{align}
    w_{kq} &= \frac{\sum_{e \in E} A_e \sum_{i<j \in e} \rho_{ijkq}^{(e)}}{C \sum_{i<j \in V} u_{ik} u_{jq}} \label{eq:w_update} \quad, \\ 
    \beta_{kz} &= \frac{\sum_{i} ( x_{iz} h_{izk} + (1 - x_{iz}) h'_{izk})}{\sum_{i, k'} ( x_{iz} h_{izk'} + (1 - x_{iz}) h'_{izk'})}\quad. \label{eq:b_update}
\end{align}
\Cref{eq:w_update} is valid when $\gamma \neq 1$ and \cref{eq:b_update} is valid when $\gamma \neq 0$. 

To obtain the updates for $u$ we distinguish two cases. In the case of $\gamma \neq 0$, differentiating \cref{eq: objective with lagrange} with respect to $u_{ik}$ yields the condition:
\begin{equation} \label{eq:u_update}
    a_{ik}\, u_{ik}^2 - (a_{ik} + b_{ik} + c_{ik})\, u_{ik} + b_{ik} = 0 \quad,
\end{equation}
where
\begin{align*}
    a_{ik} & = (1 - \gamma)\, C \sum_{j \in V, j \neq i} \sum_{q=1}^K u_{jq} w_{kq} \quad, \\
    b_{ik} &= (1 - \gamma) \sum_{e \in E: i \in e} A_e \sum_{j\neq i \in e} \sum_{q=1}^K \rho_{ijkq}^{(e)} + \gamma \sum_{z=1}^Z x_{iz} h_{izk} \quad,\\
    c_{ik} &= \gamma \sum_{z=1}^Z (1 - x_{iz}) h'_{izk} \quad .
\end{align*}
The updated values for $u_{ik}$ are found by numerically solving \cref{eq:u_update}.
We take the smallest root of \cref{eq:u_update}, as this is guaranteed to be in $(0,1)$, as we show in Supplementary Note B. This update automatically yields a value of $u_{ik}$ in $[0, 1]$, therefore the constraints on $u$ are inactive and we do not need to differentiate with respect to the Lagrange multipliers $\lambda_{ik}^{(u)}$.
 
In the case $\gamma = 0$, we differentiate \cref{eq: objective with lagrange} with respect to both $u_{ik}$ and the Lagrangian multipliers $\lambda_{ik}^{(u)}$ to obtain the update
\begin{equation}\label{eq:u_update2}
    u_{ik} = \frac{\sum_{e \in E: i \in e} A_e \sum_{j\neq i \in e} \sum_{q=1}^K \rho_{ijkq}^{(e)}}{C \sum_{j \in V, j \neq i} \sum_{q=1}^K u_{jq} w_{kq} + \lambda_{ik}^{(u)}} \, , 
\end{equation}
which is exactly the same as those of the \hymmsbm{} model \cite{ruggeri2023generalized}, except that in our case we have $\lambda_{ik}^{(u)}$ which constrains $u_{ik}\in [0,1]$. Thus, our model is as powerful as \hymmsbm{} when $\gamma = 0$, but, when the attributes correlate well with the communities, our model can utilize this information to boost performance. In practice, in the latter case, cross-validation would yield $\gamma >0$.

The EM algorithms finds a local maximum for a given starting point, which is not guaranteed to be the global maximum. Therefore, the algorithm is run several times and the best parameters are chosen based on the run that gives the highest log-likelihood.\\
A pseudocode for the algorithmic implementation is given in \cref{pseudocode:EM}. 

\begin{algorithm} 
\caption{HyCoSBM: EM algorithm} \label{pseudocode:EM}
\begin{algorithmic}
\State {\bf Inputs:} hypergraph $A$, covariates $X$, hy perparameters $\gamma$ and $K$
\State {\bf Outputs:} inferred $(u, w, \beta)$
\\
\State $u, w, \beta \gets init(u, w, \beta)$ : \texttt{Randomly initialize the parameters}
\While{\texttt{convergence not reached}}
\State $\rho, h, h' \gets update(\rho, h, h')$\Comment{\crefrange{eqn:rho}{eqn:h1}}
\State $u \gets update(u)$ \Comment{\cref{eq:u_update} or \cref{eq:u_update2}}
\If{$\gamma \neq 1$}
\State $w \gets update(w) $ \Comment{\cref{eq:w_update}}
\EndIf
\If{$\gamma \neq 0$}
\State $\beta \gets update(\beta) $ \Comment{\cref{eq:b_update}}
\EndIf
\EndWhile
\end{algorithmic}
\end{algorithm}

\section*{Hyperedge prediction and cross-validation}\label{apx:empirical}

For all experiments with real datasets we used 5-fold cross-validation with the test AUC as performance metric to select the hyperparameters $K$ and $\gamma$. We varied $K \in \ccup{2,\dots,30}$ and $\gamma \in [0.0,1.0]$. The set of hyperedges was split into 80\% and 20\% for training and testing.
The AUC is calculated by comparing the Poisson probabilities assigned to a given existing hyperedge against that of a randomly generated hyperedge of the same size. Since comparing all possible pairs of observed-unobserved edges is unfeasible, we estimate the AUC via sampling. For every observed edge in the dataset, we draw an edge of the same size uniformly at random, and compute the relative Poisson probabilities. The resulting Poisson probabilities are saved in a vector $R_1$ for the observed edges and $R_0$ for the randomly generated ones. We then compute the AUC as 
\begin{equation*}
    AUC=\frac{\sum (R_1 > R_0) + 0.5 \sum (R_1 == R_0)}{|R_1|} \quad , 
\end{equation*}
where $\sum (R_1 > R_0)$ stands for the number of times the Poisson probability of the positive hyperedge was higher than the negative one, $\sum (R_1 == R_0)$ when they were equal, and the total number $|R_1|$ of comparisons made is equal to the number of hyperedges in the test set.


\section*{Data Availability}

The data that support the findings of this study are publicly available.
The contact datasets at \url{http://www.sociopatterns.org/}; the
political interactions datasets at \url{https://www.cs.cornell.
edu/~arb/data/}; the gene-disease dataset at \cite{pinero2020disgenet}; the Enron dataset at \cite{klimt2004enron}; the New York City taxi data at \cite{nycdata}.

\section*{Code Availability}
The open source codes and executables are available at \repolink\  and at \cite{badalyan2024hycosbm}.  The code uses the \texttt{HGX} Python library \cite{lotito2023hypergraphx}.

\bibliographystyle{ScienceAdvances}
\bibliography{bibliography}

\section*{Acknowledgements}
N.R. acknowledges support from the Max Planck ETH Center for Learning Systems. C.D.B. was supported by the Cyber Valley Research Fund.

\section*{Author contributions}
A.B developed the algorithm and performed the experiments. A.B, N.R. and C.D.B. all conceived the research, analyzed the results and wrote the manuscript. 

\section*{Competing interests}
The authors declare no competing interests.

 \newpage
\appendix
 \onecolumngrid

\section{Synthetic data generation}
\label{sec supp: data generation}
We generated synthetic networks using the sampling algorithm of \hymmsbm{} \cite{ruggeri2023framework}, with implementation as in the \texttt{HGX} library described in \cite{lotito2023hypergraphx}. We set parameters as follows: $N = 500$, $|E| = 2720$ and $K = \{2, 3, 5, 10\}$.
We specify the number of hyperedges of each size using the dimension sequence \texttt{\small dim\_seq = \{2: 300, 3: 300, 4: 200, 5: 200, 6: 150, 7: 150, 8: 150, 9: 150, 10: 120, 11: 120, 12: 120, 13: 120, 14: 100, 15: 100, 16: 100, 17: 100, 18: 80, 19: 80, 20: 80\}}.

The attributes were generated to match the community structure. 
In all experiments, we set $Z=K$, and produce the attribute matrix $X$ as follows. First, the matrix $X$ is initialized equal to the community assignments $u$ of the nodes. Then, for a fraction $\rho$ of the nodes, we replace the corresponding attribute with a random one. We perform experiments with $\rho$ ranging in $\{0.1, 0.2, 0.3, 0.4, 0.5, 0.6, 0.7, 0.8, 0.9 \}$.

We randomly generated $10$ instances of higher-order networks and $10$ instances of attributes for each configuration.
The value of $\gamma$ used in these experiment was equal to the proportion of non-shuffled attributes $\gamma = 1 - \rho$.

\section{Solving for the membership matrix updates}
\label{sec supp: u update numerical}
We have the following equation to solve in order to find $u_{ik}$:
\begin{equation} \label{eq:u_update2}
    a_{ik}\, u_{ik}^2 - (a_{ik} + b_{ik} + c_{ik})\, u_{ik} + b_{ik} = 0 \quad,
\end{equation}
where $a_{ik}, b_{ik}, c_{ik}$ are all positive values. \\
This expression can be presented as a general quadratic equation 
\begin{equation}
    a x^2 - b x + c = 0 \quad,
\end{equation}
where $a, b, c$ are positive numbers and $b > a + c$.
The resulting discriminant $\Delta$ is given by
\begin{equation}
    \Delta = b^2 - 4 a c > (a + c)^2 - 4 a c = (a - c)^2 > 0\quad.
\end{equation}
Hence the discriminant is positive and there exist two distinct and real solutions to the equation. 

Now we show that the smallest root $x_0 = \frac{b - \sqrt{\Delta}}{2a}$ satisfies the constraints on $u$, that is $0 \le x_0 \le 1$. The fact that $x_0 \ge 0$ derives directly from the fact that $b \ge \sqrt{\Delta}$.
\\
Then, we show that $x_0 < 1$:
\begin{align*}
            \frac{b - \sqrt{\Delta}}{2a} &< 1 \\
    \iff    b - \sqrt{\Delta} &< 2a \\
    \iff    b^2 - 4ab + 4a^2 &< \Delta \\
    \iff    a + c &< b \quad.
\end{align*}

Similarly, it can be shown that the root $x_1 = \frac{b + \sqrt{\Delta}}{2a}$ does not yield a valid update for $u$, as 
\begin{equation*}
    \frac{b + \sqrt{\Delta}}{2a} < 1 \iff a + c > b \, ,
\end{equation*}
which is never satisfied.

\section{Alternative formulation for excluding attributes}
\label{sec supp: multinomial}
In the main manuscript we have described a model that allows a node to have multiple values for one attribute type. While in certain cases attributes could be excluding, e.g. age can take only one value, this can still be handled by that model. Alternatively, one can modify the model by assuming a probability distribution that explicitly imposes the choice of only one value, e.g. with a Multinomial distribution. Here we illustrate how our model can be adapted to this case and highlight the main differences with the formulation adopted in the main manuscript.

We assume that each entry $X_{iz}$ is extracted from a Multinomial distribution with parameter $\pi_{iz}=\sum_{k=1}^K u_{ik}\, \beta_{kz}$. Then, the likelihood of the attributes matrix can be modelled as:
\begin{equation}
    P_X(X|U, \beta) = \prod_{i \in V} Mult \left(X_i; \pi_{i} \right) \quad,
\end{equation}
which gives the following log-likelihood for the attributes:
\begin{equation}
    L_X(U, \beta) = \sum_{i=1}^N \sum_{z=1}^Z x_{iz} \log(\pi_{iz}) = \sum_{i=1}^N \sum_{z=1}^Z x_{iz} \log \left( \sum_{k=1}^K \beta_{kz} u_{ik} \right) .
\end{equation}
Using a standard variational approach to lower bound the log-likelihood we get:
\begin{equation} \label{eq:likelihood multinomial}
    \mathcal{L}_X(U, \beta, h) = \sum_{i, z, k} x_{iz} \left[ h_{izk} \log (\beta_{kz} u_{ik}) - h_{izk} \log(h_{izk}) \right]
\end{equation}
with the equality reached when 
\begin{equation}
    h_{izk} = \frac{\beta_{kz} u_{ik}}{\sum_{k'} \beta_{k'z} u_{k'z}} .
\end{equation}

While \cref{eq:likelihood multinomial} looks simpler than the one we derived using the Bernoulli distribution, the main challenge is
to obtain a tractable solution for the updates of $u_{ik}$ as we introduce constraints on the parameters.
The constraint for $\beta$ is $\sum_z\, \beta_{kz}=1,\ \forall k $. However, now the constraint on $u_i$ involves all of the entries of this vector at the same time: $\sum_k \, u_{ik}=1$, increasing the complexity of the subsequent derivations. In addition, we still impose positivity $u_{ik}\geq 0$, $\forall i,k$.\\
By introducing Lagrange multipliers $\lambda = (\lambda^{(\beta)}, \lambda^{(u)}, \mu^{(u)} )$, where $\lambda^{(u)}$ controls the summation to one term and $\mu^{(u)}$ positivity, we get the following update for $u_{ik}$:


\begin{align} \label{eq:u_update multinomial}
    u_{ik} = \frac
    {(1 - \gamma) \sum_{e \in E: i \in e} A_e \sum_{j\neq i \in e} \sum_{q} \rho_{ijkq}^{(e)}  + \gamma \sum_z x_{iz} h_{izk}}
    {\lambda_i^{(u)} - \mu_{ik}^{(u)} + (1 - \gamma) C \sum_{j \in V, j \neq i} \sum_{q=1}^K u_{jq} w_{kq}} \quad.
\end{align}
To estimate $\lambda^{(u)}$ we need to solve the following equation:
\begin{align} \label{eq:root_finding}
    \sum_{k = 1}^K u_{ik} = \sum_{k = 1}^K \frac
    {(1 - \gamma) \sum_{e \in E: i \in e} A_e \sum_{j\neq i \in e} \sum_{q} \rho_{ijkq}^{(e)}  + \gamma \sum_z x_{iz} h_{izk}}
    {\lambda_i^{(u)} - \mu_{ik}^{(u)} + (1 - \gamma) C \sum_{j \in V, j \neq i} \sum_{q=1}^K u_{jq} w_{kq}} = 1 \quad.
\end{align}

\Cref{eq:root_finding} cannot be solved in closed-form but can be solved numerically, e.g. with root-finding methods. However, this can slow down the implementation considerably and may not always converge to a solution. 

\section{The advantages of using a hypergraph representation}
To demonstrate possible advantages of utilizing a hypergraph representation, and specifically of enriching it with node attributes, we compare the performance of \hycosbm\ against those of a dyadic representation of a hypergraph on a hyperedge prediction task. 

There are various ways that one can use to project an hypergraph into a standard network structure with pairwise edges. Here we consider the clique expansion, where for each hyperedge one creates a clique with all the possible pairs of nodes in it. The resulting network is the union of these cliques and edge weights are the numbers of hyperedges in which a pair of nodes was contained in. This is a popular approach when investigating hypergraphs, see for example \cite{chodrow2021generative,contisciani2022inference}.
We then apply an algorithm that has similar characteristics as \hycosbm\, but is only valid in networks.  
As the focus of this work is on utilyzing node attributes as additional information, 
as a comparison we use MTCOV \cite{contisciani2020community}, a probabilistic model for networks that is able to utilize node attributes to efficiently infer the network structure. Similarly to \hycosbm, it also uses latent variables like community memberships.
As approaches for (pairwise) networks only output the probability of observing pairwise interactions, we define the probability of a hyperedge as the product of all edges belonging to its clique expansion. We refer to this approach of using MTCOV on the clique expansion as \clique.

\begin{table}[hptb]
\centering
\setlength{\tabcolsep}{3pt}
\setlength{\arrayrulewidth}{0.1pt}
\begin{tabular}{lrrrr}
	\toprule
	Dataset & $N$ & $|E|$ & $|E_2|$ & $|E_{\text{Clique-Exp}}|$ \\
	\midrule
	High School  & 327 & 7818 &5498& 5818 \\
	Hospital  & 75 & 1825 &1108 &1139 \\
	Primary School  & 242 & 12704 &7748& 8317 \\
	Workplace  & 92 & 788 & 742& 755 \\
	Gene Disease  & 9262 & 3128 & 886& 2837026 \\
	NYC taxi trips Mon-Tue 17-19 & 214 & 523 & 64&18568 \\
	NYC taxi trips Sat-Sun 00-02 & 214 & 476 &53&  16146 \\
	\bottomrule
\end{tabular}
    \caption{\textbf{Statistics on graphs obtained by clique expansion}. Number of nodes $N$, number of  hyperedges $|E|$,  number of hyperedges $|E_2|$ of size $2$, and number of (dyadic) edges $|E_{\text{Clique-Exp}}|$ obtained by clique expansion are reported. The latter three quantities consider the number of unique edges, not accounting for edge weights.}
\label{tab:clique_exp_stats}
\end{table}

As a preliminary analysis, in \cref{tab:clique_exp_stats} we compare the number of interactions observed in different real-world datasets and their relative clique expansion.
In contacts datasets, we observe that the majority of interactions are pairwise, with the bulk of the interactions being of sizes two and three. In addition, many higher-order edges overlap, as they contain pairs of nodes that are already present in other hyperedges. As a result, the number of unique hyperedges $|E|$ is larger than the number of unique pairwise edges in the clique expansion $|E_{\text{Clique-Exp}}|$.

This could be a reason for not observing a significant difference between \clique\ and \hycosbm\ on the High School and Primary School datasets in predicting hyperedges. Nevertheless, it is difficult to draw a general conclusion as there are several variables that could contribute to prediction performance (e.g. how hyperedges overlap by sharing subset of nodes, etc...). For instance, in other datasets similar to the contacts in schools, \hycosbm\ outperforms \clique; this happens in Hospital with AUC equal to $0.776$ versus $0.714$ and in Workplace with respective AUC scores of $0.81$ and $0.774$, as reported in \cref{tab:hye_rep_results}.  

\begin{table}[hptb]
	\centering
	\setlength{\tabcolsep}{3pt}
	\setlength{\arrayrulewidth}{0.1pt}
\begin{tabular}{llllllcclclcc}
	\toprule
	Dataset & Attribute & $Z$ & \multicolumn{3}{c}{\hycosbm} & \multicolumn{2}{c}{\hymmsbm} & \multicolumn{3}{c}{\clique} \\
	& & & $K$ & $\gamma$ & AUC & $K$ &   AUC & $K$ & $\gamma$ & AUC  \\
	\midrule
	Gene Disease & DPI & 25 & 30 & 0.500 & $0.9 \pm 0.07$ & 2 & $0.84 \pm 0.122$ & 5 & 0.995 & $0.682 \pm 0.015$ \\
	\cline{1-11} 
	\multirow[t]{4}{*}{High School} & class &  9 & 11 & 0.995 & $0.899 \pm 0.011$ &  \multirow[c]{4}{*}{24} & \multirow[c]{4}{*}{$0.884 \pm 0.006$}& 24 & 0.995 & $0.906 \pm 0.008$ \\
	& has filled questionnaire &  2 & 21 & 0.800 & $0.892 \pm 0.013$ &  & & 29 & 0.200 & $0.894 \pm 0.007 $  \\
	& has facebook &  2 & 15 & 0.950 & $0.888 \pm 0.008$ &  & & 30 & 0.800 & $0.892 \pm 0.013$\\
	& sex & 2 & 16 & 0.800 & $0.889 \pm 0.009$ & & & 25 & 0.600 & $0.895 \pm 0.009$
	\\
	\cline{1-11}
	\multirow[t]{2}{*}{Primary School} & class & 11 & 10 & 0.600 & $0.841 \pm 0.013$ & \multirow[c]{2}{*}{11} & \multirow[c]{2}{*}{$0.841 \pm 0.007$} & 24 & 0.995 & $0.847 \pm 0.010$\\
	& sex & 2 & 12 & 0.200 & $0.841 \pm 0.007$ & & & 23 & 0.100 & $0.836 \pm 0.007$\\
	\cline{1-11}
	Hospital & status & 4 & 2 & 0.200 & $0.776 \pm 0.032$ & 2 & $0.758 \pm 0.016$ & 23 & 0.995 & $0.714 \pm 0.046 $ \\
	\cline{1-11}
	Workplace & department & 5 & 5 & 0.995 & $0.81 \pm 0.02$ & 5 & $0.752 \pm 0.039$ & 6 & 0.990 & $0.774 \pm 0.025 $ \\
	\bottomrule
\end{tabular}
	\caption{\textbf{AUC scores achieved by \hycosbm, \hymmsbm, \clique}. The best results achieved by 5-fold cross validation as well as best $\gamma$ and $K$ by all models are reported. For the \clique on Gene Disease dataset, the maximum number of communities used during cross-validation was $K=7$ due to computational constraints. AUC values and errors are averages and standard deviations over 5 cross-validation folds. }
	\label{tab:hye_rep_results}
\end{table}

On the other hand, the clique expansion obtained from Gene Disease contains about 2.8 million dyadic edges, compared to only 3128 hyperedges.  This is because it contains many hyperedges of large sizes (also of size $\sim1000$ nodes). This makes it difficult to run a code on the clique expansion, even when the complexity is only linear in the number of edges, as it is the case for MTCOV. We were able to run the cross-validation procedure only for small values of $K \leq 7$, which results in poor performance of the model, compared to both \hycosbm\ and \hymmsbm. This shows that clique expansions could be significantly limiting, as one may not even be able to run a standard network model on them due to computational challenges. This is particularly the case in hypergraphs with large hyperedge sizes. On the contrary, both  \hycosbm\ and \hymmsbm\ are not significantly impacted by this and are efficient to run on large and sparse hypergraphs.

\section{Additional results of community detection}
\begin{figure*}[t]
    \centering
    \includegraphics[width=1\textwidth]{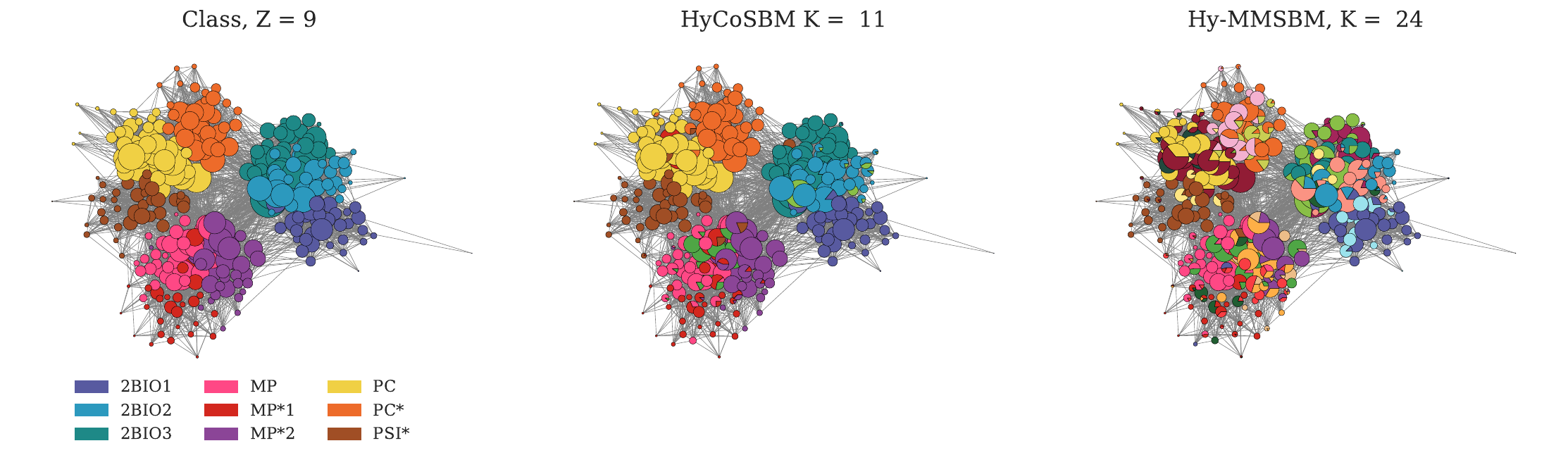}
    \caption{Communities detected in a High School dataset of close-proximity interactions. We give the whole dataset as input to the algorithms, and compare the inferred communities against the \texttt{class} attribute (top left). The plot shows that both \hycosbm{} and \hymmsbm{} detect communities aligned with the attribute, but with a number of communities greater than the number of attribute values. AUC values are slightly higher for \hycosbm, see Table III in the main manuscript.}
    \label{fig:high-school}
\end{figure*}

\begin{figure*}[t]
    \centering
    \includegraphics[width=1\textwidth]{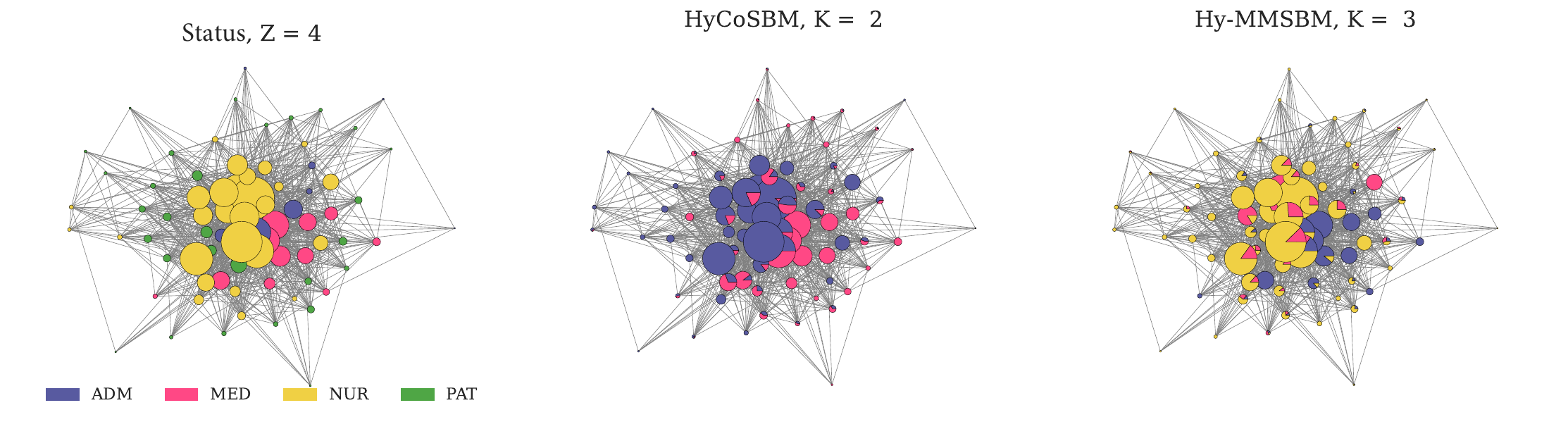}
       \caption{Communities detected in Hospital dataset using $60\%$ of hyperedges. We give in input to the algorithms $60\%$ of hyperedges and compare the inferred communities against the attribute \texttt{status} (NUR=paramedical staff; PAT=Patient; MED=Medical doctor; ADM=administrative staff) (top left). This plot shows that both \hycosbm\ and \hymmsbm{} detect fewer communities than the division indicated by attributes, with \hycosbm{} achieving a higher AUC that \hymmsbm{}, see Fig. 2 in the main manuscript.}
    \label{fig:hospital}
\end{figure*}

We provide additional results about communities detected in the High School and Hospital datasets in \cref{fig:high-school,fig:hospital}.

\end{document}